\documentstyle[multicol,aps,prl,psfig,array]{revtex} 

\renewcommand{\narrowtext}{\begin{multicols}{2} \global\columnwidth20.5pc}
\renewcommand{\widetext}{\end{multicols} \global\columnwidth42.5pc}
\def\be{\beta}
\def\ga{\gamma}
\def\de{\delta}
\def\ep{\epsilon}

\def\et{\eta}
\def\th{\theta}

\def\la{\lambda}

\def\ph{\phi}

\def\ch{\chi}

\def\om{\omega}
\def\Ga{\Gamma}
\def\De{\Delta}

\def\La{\Lambda}

\def\Ps{\Psi}
\def\Om{\Omega}

\def\cA{{\cal A}}

\def\cN{{\cal N}}

\def\fr#1#2{{{#1} \over {#2}}}
\def\frac#1#2{\textstyle{{{#1} \over {#2}}}}

\def\prt{\partial}

\def\ket#1{|{#1}\rangle}
\def\bra#1{\langle{#1}|}

\def\half{{\textstyle{1\over 2}}}
\def\lsim{\mathrel{\rlap{\lower4pt\hbox{\hskip1pt$\sim$}}
    \raise1pt\hbox{$<$}}}
\def\gsim{\mathrel{\rlap{\lower4pt\hbox{\hskip1pt$\sim$}}
    \raise1pt\hbox{$>$}}}

\def\X{\hat X}
\def\Y{\hat Y}
\def\Z{\hat Z}
\def\x{\hat x}
\def\y{\hat y}
\def\z{\hat z}

\def\Re{\hbox{Re}\,}
\def\Im{\hbox{Im}\,}

\def\ol#1{\overline{#1}}
\newcommand{\beq}{\begin{equation}}
\newcommand{\eeq}{\end{equation}}
\newcommand{\bea}{\begin{eqnarray}}
\newcommand{\eea}{\end{eqnarray}}
\newcommand{\rf}[1]{(\ref{#1})}

\begin{document}

\title{ CPT, T, and Lorentz Violation
in Neutral-Meson Oscillations}    
\author{V.\ Alan Kosteleck\'y}
\address{Physics Department, Indiana University, 
         Bloomington, IN 47405, U.S.A.}
\date{IUHET 430, March 2001; Phys.\ Rev.\ D, in press} 

\maketitle

\begin{abstract}
Tests of CPT and Lorentz symmetry using neutral-meson oscillations
are studied within a formalism 
that allows for indirect CPT and T violation of arbitrary size
and is independent of phase conventions.
The analysis is particularly appropriate
for studies of CPT and T violation 
in oscillations of the heavy neutral mesons $D$, $B_d$, and $B_s$.
The general Lorentz- and CPT-breaking standard-model extension
is used to derive an expression for the parameter for CPT violation. 
It varies in a prescribed way
with the magnitude and orientation of the meson momentum
and consequently also with sidereal time.
Decay probabilities are presented
for both uncorrelated and correlated mesons,
and some implications for experiments are discussed.

\end{abstract}


\narrowtext

\section{Introduction}
\label{intro}

The original discovery of CP violation in the neutral-kaon system
\cite{ccft}
has led to numerous theoretical and experimental studies 
of discrete symmetries in neutral-meson oscillations
\cite{ww}.
Much of the effort has been focused on the $K$ system,
but the advent of high-statistics experiments
involving the heavy neutral mesons, 
in particular the $B_d$ mesons
\cite{cbb},
has opened the door for a broader class of investigations.

In a neutral-meson system,
the violation of CP symmetry includes the possibility 
of CPT violation
\cite{sachs,cpt98}.
For the $K$ system,
CPT violation in oscillations 
can be parametrized by a complex quantity $\de_K$
that is known to be small or zero
\cite{lw}.
Under the \it ad hoc \rm assumption that $\de_K$ is
a constant complex number,
experiments have established that its real and imaginary parts 
are no greater than about $10^{-4}$
\cite{e773,cplear}.

The assumption of constant nonzero $\de_K$ is known to fail 
in conventional quantum field theory.
The nature of $\de_K$ is determined by 
the properties of the theory under Lorentz transformations.
For any realistic Lorentz-invariant quantum field theory
such as the standard model,
the CPT theorem shows that $\de_K$ must be zero 
\cite{sachs}.
If instead Lorentz violation is allowed,
then using an explicit and general standard-model extension 
\cite{ck}
to calculate $\de_K$ reveals 
that it varies with the meson 4-momentum
\cite{ak,ak2}.
This variation has recently been exploited by the KTeV Collaboration
in placing a qualitatively new bound on 
CPT violation in the neutral-$K$ system
\cite{k99}. 

For systems involving the heavy mesons $D$, $B_d$, $B_s$
several CPT tests have been proposed
\cite{kp,ckvk,msas},
and bounds have been obtained
in some recent experiments with the $B_d$ system
\cite{bexpt}.
All these results rely on the assumption
of a nonzero constant complex parameter for CPT violation.
However,
as in the $K$ system,
this assumption fails in realistic quantum field theories:
either the parameter vanishes by the CPT theorem,
or it depends on the 4-momentum of the meson.

The present work provides a general treatment 
of CPT violation in neutral-meson oscillations
in the context of quantum field theory
allowing for Lorentz violation.
A convenient formalism is adopted 
that is independent of phase conventions
and allows for CPT and T violation of arbitrary size
in any neutral-meson system.
The complex parameter for CPT violation
is calculated in the general Lorentz-violating standard-model extension,
revealing a well-defined variation 
with the magnitude and orientation of the meson momentum
and a corresponding variation with sidereal time.
Some experimentally relevant 
decay probabilities and asymmetries are derived
for both uncorrelated and correlated mesons.
The results obtained here complement the analyses in earlier works,
which described some essential physics 
\cite{ak}
and obtained expressions valid for small CPT violation
in the $K$, $D$, $B_d$, and $B_s$ systems
\cite{ak2}.

Section \ref{basics} provides background information
and fixes some notational conventions.
A suitable parametrization of the effective hamiltonian
for the time evolution of a neutral-meson state
with CPT and T violation of arbitrary size
is presented in section \ref{formalism}.
The calculation of the complex parameter for CPT violation
is given in section \ref{theory}.
Implications for experiment are considered
in section \ref{expt}.
The appendix contains a brief description
of other formalisms adopted in the literature.
Throughout this work,
a strong-interaction eigenstate is denoted generically by $P^0$,
where $P^0$ is one of $K^0$, $D^0$, $B_d^0$, $B_s^0$,
and the corresponding opposite-flavor antiparticle 
is denoted $\overline{P^0}$.

\section{Basics}
\label{basics}

An arbitrary neutral-meson state is 
a linear combination of the Schr\"odinger wave functions 
for the meson $P^0$ and its antimeson $\overline{P^0}$.
This combination can be represented as a two-component object $\Ps(t)$,
with time evolution governed by 
a 2$\times$2 effective hamiltonian $\La$ 
according to the Schr\"odinger-type equation 
\cite{lw}
\beq
i\prt_t \Ps = \La \Ps.
\label{seq}
\eeq
Throughout this paper,
subscripts $P$ are understood on $\Ps$,
on the components of the effective hamiltonian $\La$,
and on related quantities 
such as meson masses and lifetimes.

The physical propagating states are the eigenstates of $\La$,
analogous to the normal modes 
of a classical two-dimensional oscillator
\cite{osc}.
In this work,
these states are generically denoted as $\ket{P_a}$ and $\ket{P_b}$.
They evolve in time as
\bea
\ket{P_a(t)}&=&\exp (-i\la_at) \ket{P_a},
\nonumber\\
\ket{P_b(t)}&=&\exp (-i\la_bt) \ket{P_b}.
\label{timevol}
\eea
The complex parameters $\la_a$, $\la_b$ are the eigenvalues of $\La$.
They can be decomposed as 
\beq
\la_a \equiv m_a - \half i \ga_a, \quad 
\la_b \equiv m_b - \half i \ga_b,
\label{mga}
\eeq
where $m_a$, $m_b$ are the propagating masses
and $\ga_a$, $\ga_b$ are the associated decay rates.
For the $K$ system,
contact with the standard notation can be made via 
the identification $m_a = m_S$, $m_b = m_L$,
$\ga_a = \ga_S$, $\ga_b = \ga_L$.
For the $D$ system,
there is no well established convention
and I use the notation in Eq.\ \rf{mga}.
For the $B_d$ and $B_s$ systems,
the relation to the standard notation can be taken as 
$m_a = m_L$, $m_b = m_H$,
$\ga_a = \Ga_L$, $\ga_b = \Ga_H$.

For calculational purposes,
it is useful to introduce a separate notation for the
sums and differences of these parameters:
\bea
\la &\equiv &\la_a + \la_b = m - \half i \ga,
\nonumber\\
\De \la &\equiv &\la_a - \la_b = - \De m - \half i \De \ga,
\label{ldl}
\eea
where $m = m_a + m_b$, $\De m = m_b - m_a$,
$\ga = \ga_a + \ga_b$, $\De \ga = \ga_a - \ga_b$.
Note in particular the choice of sign in the
definition of $\De\ga$,
which coincides with that in the $K$ system
but is the negative of the quantity $\De\Ga$
often adopted in the $B_d$ system.
The reader can therefore make direct contact 
with results in the latter convention 
by identifying $\De\ga \equiv - \De\Ga$
in any equation in this work.

The off-diagonal components of $\La$
control the flavor oscillations between $P^0$ and $\overline{P^0}$.
Indirect CPT violation occurs
if and only if the difference of diagonal elements of $\La$
is nonzero, 
$\La_{11} - \La_{22} \neq 0$.
Indirect T violation occurs 
if and only if the magnitude of the ratio of 
off-diagonal components of $\La$ differs from 1, 
$|\La_{21}/\La_{12}|\neq 1$.

\it A priori, \rm
the effective hamiltonian $\La$ 
can be parametrized by eight independent real quantities.
Four of these can be specified 
in terms of the masses and decay rates,
two describe CPT violation, 
and one describes T violation.
The remaining parameter,
determined by the relative phase between
the off-diagonal components of $\La$,
is physically irrelevant.
It can be dialed at will
by rotating the phases of the $P^0$ and $\overline{P^0}$ wave functions
by equal and opposite amounts.
The freedom to perform such rotations exists
because the wave functions are eigenstates of the strong interactions,
which preserve strangeness, charm, and beauty.
Under a rotation of this type involving a phase factor
of $\exp(i\ch)$ for the $P^0$ wave function,
the off-diagonal elements of $\La$ are multiplied by
equal and opposite phases,
becoming $\exp(2 i \ch)\La_{12}$ and $\exp(-2i\ch)\La_{21}$.

\section{Formalism}
\label{formalism}

Since relatively little experimental information is 
available about CPT and T violation 
in the heavy neutral-meson systems,
a general parametrization of $\La$ is appropriate.
It is desirable to have a parametrization 
that is model independent,
valid for arbitrary size CPT and T violation,
independent of phase conventions,
and expressed in terms of mass and decay rates insofar as possible.
A parametrization of this type was originally 
introduced by Lavoura in the context of the kaon system
\cite{ll,lsx}.
For simplicity,
it is also attractive to arrange matters 
so that the quantities controlling T and CPT violation
are denoted by single symbols
that are distinct from other frequently used notation.
In this section,
a parametrization convenient
to the four meson systems
and satisfying all the above criteria
is presented and related to formalisms 
often used in the literature. 

For a complex $2\times 2$ matrix,
it is possible to write the two diagonal elements
as the sum and difference of two complex numbers.
It is also possible to write the off-diagonal elements
as the product and ratio of two complex numbers.
Using these two facts,
which ultimately permit the clean representation
of T- and CPT-violating quantities, 
a general expression for $\La$ can be taken as:
\beq
\La = 
\half \De\la
\left( \begin{array}{lr}
U + \xi 
& 
\quad VW^{-1} 
\\ & \\
VW \quad 
& 
U - \xi 
\end{array}
\right),
\label{uvwx}
\eeq
where the parameters $UVW\xi$ are complex.
The factor $\De\la/2$ has been extracted from $\La$
to make these parameters dimensionless
and to avoid factors of 2 in expressions below.

The requirements that the trace of the matrix is tr$~\La = \la$ 
and that the determinant is 
$\det \La = \la_a \la_b$
impose the identifications 
\beq
U \equiv \la/\De\la, \quad 
V \equiv \sqrt{1 - \xi^2}
\label{uvdef}
\eeq
on the complex parameters $U$ and $V$.
The free parameters in Eq.\ \rf{uvwx} are therefore 
$W$ and $\xi$.
These can be regarded as four independent real quantities:
$W = w \exp (i\om)$,
$\xi = \Re\xi + i \Im \xi$.
One of these four real numbers,
the argument $\om$ of $W$,
is arbitrary and physically irrelevant.
It changes under the phase redefinitions discussed 
at the end of the previous section.
The other three are physical.
The modulus $w$ of $W$ controls T violation,
with $w = 1$ if and only if T is preserved
\cite{fn00}.
The two remaining real numbers, 
$\Re\xi$ and $\Im\xi$,
control CPT violation
and both are zero if and only if CPT is preserved.
The quantities $w$ and $\xi$ 
can be expressed in terms of the components of $\La$ as
\cite{fn0}
\beq
w = \sqrt{|\La_{21}/\La_{12}|},
\quad
\xi = \De\La/\De\la,
\label{wxiexpr}
\eeq
where $\De\La = \La_{11}-\La_{22}$.

In this $w\xi$ formalism,
the three parameters for CP violation $w$, $\Re\xi$, $\Im\xi$
are dimensionless and independent of phase conventions.
They are phenomenologically introduced
and therefore are independent of specific models.
However,
this does \it not \rm imply that they are 
necessarily constant numbers.
Indeed,
the assumption of constancy for $\xi$
frequently made in the literature
is a special choice that strongly restricts 
the generality of the parametrization
and which according to the CPT theorem is inconsistent 
with the fundamental structure of Lorentz-invariant quantum field theory.
In fact,
if the requirement of exact Lorentz symmetry is relaxed,
then $\xi$ cannot be a constant quantity 
within the framework of quantum field theory 
but instead must vary with the momentum of the meson.
Since CPT violation is a profound effect, 
it is unsurprising that the parameter $\xi$
has features different from $w$.
The choice of the notation $\xi$ (rather than, say, $X$)
in Eq.\ \rf{uvwx} 
has been made to emphasize this crucial fact.

The physical states with definite mass and lifetimes
are the eigenstates of $\La$.
In the $w\xi$ formalism,
they take the form 
\bea
\ket{P_a} &=&
\cN_a (\ket{P^0} + A \ket{\overline{P^0}}) , 
\nonumber \\
\ket{P_b} &=&
\cN_b (\ket{P^0} + B \ket{\overline{P^0}}) ,
\label{statesdef}
\eea         
where
\beq
A = (1 - \xi) W/V, \quad
B = -(1 + \xi) W/V.
\eeq
The normalizations $\cN_a$, $\cN_b$ in Eq.\ \rf{statesdef}
can be chosen as desired.
For unit-normalized states,
the normalizations are 
\bea
\cN_a &=& \exp(i \et_a)/\sqrt{1 + |A|^2}, 
\nonumber\\
\cN_b &=& \exp(i \et_b)/\sqrt{1 + |B|^2}, 
\label{norms}
\eea
where $\et_a$ and $\et_b$ are phases
that can be chosen freely.
For the analysis of physical observables 
in the following sections,
the values of these phases are irrelevant
\cite{fn1}.

\widetext

\medskip

\halign{
\indent\qquad\hfil#\hfil 
&\qquad\hfil#\hfil 
&\qquad\hfil#\hfil 
&\qquad\hfil#\hfil 
&\qquad\hfil#\hfil 
\cr
\noalign{\medskip}
\hline
\noalign{\medskip}
Formalism
&
Parameters depend
&
$\la$, $\De\la$
&
CPT parameter
&
T parameter
\cr
&
on phase convention?
&
given as 
&
(complex)
&
(real)
\cr
\noalign{\smallskip}
\hline
\noalign{\medskip}
$w \xi $
&
No
&
$\la$, $\De\la$
&
$\xi$
&
$w$
\cr
\noalign{\medskip}
$M\Ga$
&
Yes
($M_{12}$, $\Ga_{12}$)
&
See Eq.\ \rf{mgamess}
&
$(M_{11} - M_{22})$
&
$\fr{| M_{12}^* - i \Ga_{12}^*/2 |}{| M_{12} - i \Ga_{12}/2 |}$
\cr
\noalign{\vskip -4pt}
&
&
&
\quad $- \half i (\Ga_{11} - \Ga_{22})$
&
\cr
\noalign{\medskip}
$D E_1 E_2 E_3$
&
Yes
($E_1$, $E_2$)
&
\hbox{\hskip-30pt}
$- 2 iD$,
$2\sqrt{E_1^2+E_2^2+E_3^2}$
&
$E_3$
&
$i(E_1 E_2^* - E_1^* E_2)$
\cr
\noalign{\medskip}
$D E \th\ph$
&
Yes 
($\ph$)
&
$-2i D$, $2E$
&
$\cos\th$
&
$|\exp(i\ph)|$
\cr
\noalign{\medskip}
$pqrs$
&
Yes
($p$, $q$, $r$, $s$)
&
$\la$, $\De\la$
&
$(ps-qr)$
&
$|pr/qs|$
\cr
\noalign{\medskip}
$\ep\de$
&
Yes
($\ep$, $\de$)
&
$\la$, $\De\la$
&
$\de$
&
$\Re\ep$, if ${\rm CP}{\hskip-10pt \Bigl/}\hskip +3pt$ small
\cr
\noalign{\smallskip}
\hline
\noalign{\smallskip}
}
\medskip
\centerline{\small
Table 1: Comparison of formalisms for neutral-meson mixing.
} 
\smallskip

\narrowtext

Some insight into the advantages of the $w\xi$ formalism 
can be obtained by comparing it to alternative formalisms 
available in the literature.
The appendix summarizes some of the more popular ones,
and Table 1 provides a comparative synopsis of their features.
The first column identifies the formalism
through the standard notation for its parameters.
The second column indicates 
the phase-convention dependence of its parameters.
The third column lists the connection between the
physical quantities $\la$, $\De\la$ 
and their expression in the given formalism.
The fourth column specifies the
complex combination of parameters that governs CPT violation
in the specified formalism,
while the last column gives the real number
controlling T violation.
Note that the final entry on the last line 
holds only for small CPT and T violation
and assumes a phase convention with $\Im \ep = 0$.

Exact relationships exist between the $w\xi$ formalism
and the other formalisms listed in Table 1,
but they can be involved and may change 
with the choice of phase conventions.
Expressing the complex parameter $\xi$ for CPT violation 
in the other parametrizations gives
\bea
\xi 
&=& 
\half \bigl[
(M_{11} - M_{22}) - \half i (\Ga_{11} - \Ga_{22})
\bigr]
\nonumber\\
&&\times
\bigl[
(M_{12} - \half i \Ga_{12}) 
(M^*_{12} - \half i \Ga^*_{12}) 
\nonumber\\
&&\qquad
+\frac 14[ (M_{11} - M_{22}) - \half i (\Ga_{11} - \Ga_{22})]^2 
\bigr]^{-1/2}
\nonumber\\
&=& 
\fr
{E_3}
{\sqrt{E_1^2+E_2^2+E_3^2}}
\nonumber\\
&=& 
\cos\th
\nonumber\\
&=& 
\fr
{(ps-qr)}
{(ps+qr)}
\nonumber\\
&\approx& 
2\de.
\label{cpt}
\eea
The last line is valid only for small $\de$ and $\ep$
and only in a special phase convention,
but shows that $\xi$ can be identified with $2\de$
for an appropriate choice of phase convention in the $K$ system.
In any case,
for the $D$, $B_d$, and $B_s$ systems,
$\xi$ appears simpler to use than $\de$
or any of the other parametrizations.

A similar exercise for the real parameter $w$ for T violation yields
\bea
w 
&=& 
|(M_{12}^* - \half i \Ga_{12}^*)/(M_{12} - \half i \Ga_{12})|^{1/2}
\nonumber\\
&=& 
|(E_1 + i E_2)/(E_1 - i E_2)|^{1/2}
\nonumber\\
&=& 
|\exp(i\ph)|
\nonumber\\
&=& 
\sqrt{|qs/pr|}
\nonumber\\
&\approx& 
1 - 2\Re\ep.
\label{t}
\eea
The last line is again valid only for small $\de$ and $\ep$
and only in a special phase convention.

The above equations reveal that the $w\xi$ formalism
is most closely related to the $DE\th\ph$ formalism,
but offers a more direct link to $\la$, $\De\la$,
an abbreviated notation for CPT violation,
and a single symbol for the 
phase-independent physical parameter for T violation.
On the more practical side,
the use of $\xi$ also avoids confusion 
with the standard use of the track orientation angles $\th$, $\ph$ 
for the meson in the detector,
which is a useful asset in the presence of 
orientation-dependent CPT-violating effects.
Overall, 
advantages of the $w\xi$ formalism include
its model independence,
its use of mass and decay rates as physical parameters,
its validity for arbitrary-size CPT and T violation,
and its independence of phase conventions.
In the present work,
use of the $w\xi$ formalism simplifies the results 
of the study of CPT violation.

\section{Theory for CPT Violation}
\label{theory}

The CPT theorem guarantees CPT invariance 
of Lorentz-symmetric quantum field theories,
including the usual standard model of particle physics.
To construct a description of CPT violation
viable at the level of quantum field theory,
it is therefore of interest to consider 
the possibility of small violations of Lorentz invariance.
A general standard-model extension allowing for
Lorentz and CPT violation is known  
\cite{ck}.
It could emerge,
for example,
as the low-energy limit of a fundamental theory at the Planck scale
\cite{kps}.
This standard-model extension provides
a quantitative microscopic theory for Lorentz and CPT violation
that is applicable to a wide class of experiments
in addition to the studies of neutral-meson oscillations
considered in the present work.
Among these are,
for example,
comparative tests of QED 
in Penning traps \cite{bkr,gg,hd,rm},
spectroscopy of hydrogen and antihydrogen \cite{bkr2,dp},
measurements of muon properties
\cite{bkl,vh},
clock-comparison experiments
\cite{ccexpt,kla,lh,db},
observations of the behavior of a spin-polarized torsion pendulum
\cite{bk,bh},
measurements of cosmological birefringence \cite{cfj,ck,jk,pvc},
and observations of the baryon asymmetry \cite{bckp}.
However,
none of these tests 
are sensitive to the sector of the standard-model extension
involved in the experiments with neutral-meson oscillations,
essentially because the latter are flavor changing
\cite{ak}. 

Using the general standard-model extension,
a perturbative calculation can be performed to 
obtain the leading-order CPT-violating contributions to $\La$.
These emerge as the expectation values of interaction terms
in the standard-model hamiltonian
\cite{kp}.
The CPT-unperturbed wave functions
$\ket{P^0}$ and $\ket{\overline{P^0}}$
are the appropriate states for constructing the expectation values.
The hermiticity of the perturbation hamiltonian 
ensures reality of the dominant contributions 
to the difference $\De\La =\La_{11} - \La_{22}$
of the diagonal terms of $\La$
and therefore constrains the form of $\La$.
It can be shown that
\cite{ak} 
\beq
\De\La \approx \be^\mu \De a_\mu
\quad ,
\label{dem}
\eeq
where $\be^\mu = \ga (1, \vec \be )$ is the four-velocity
of the meson state in the observer frame.
The effect of Lorentz and CPT violation 
in the standard-model extension 
appears in Eq.\ \rf{dem} via the factor 
$\De a_\mu = r_{q_1}a^{q_1}_\mu - r_{q_2}a^{q_2}_\mu$,
where $a^{q_1}_\mu$, $a^{q_2}_\mu$
are CPT- and Lorentz-violating coupling coefficients
for the two valence quarks in the $P^0$ meson,
and where $r_{q_1}$ and $r_{q_2}$
are quantities resulting from quark-binding 
and normalization effects 
\cite{kp}.
The coefficients  
$a^{q_1}_\mu$, $a^{q_2}_\mu$
for Lorentz and CPT violation
have mass dimension one
and emerge from terms in the lagrangian for the standard-model extension
of the form $- a^q_\mu \overline{q} \ga^\mu q$,
where $q$ specifies the quark flavor.

The 4-velocity and hence 4-momentum dependence in Eq.\ \rf{dem} 
confirms the failure of the usual assumption of a constant  
parameter for CPT violation.
This dependence has substantial implications for experiments,
since CPT observables will typically vary with the momentum magnitude 
and orientation of the mesons.
As a result,
the CPT reach of an experiment is affected by 
the meson momentum spectrum and angular distribution
\cite{ak,ak2}.

A significant consequence of the 4-momentum dependence
arises from the rotation of the Earth 
relative to the constant vector $\De\vec a$.
This leads to sidereal variations in some observables
\cite{ak,ak2}.
The point is that the analysis leading to Eq.\ \rf{dem}
is performed in the laboratory frame,
which rotates with the Earth.
The resulting sidereal time dependence
can be exhibited explicitly by
converting the expression for $\De\La$
to a nonrotating frame.

Denote the spatial basis in the nonrotating frame by 
$(\X,\Y,\Z)$
and that in the laboratory frame by 
$(\x,\y,\z)$.
Following Ref.\ \cite{kla},
define the nonrotating-frame basis $(\X,\Y,\Z)$ 
to be compatible with celestial equatorial coordinates
\cite{celestial}
with $\Z$ aligned along the Earth's rotation axis.
The $\z$ axis in the laboratory frame 
can be chosen for maximal convenience.
For collimated mesons,
it may be useful to take it as the beam direction.
In a collider, 
the direction of the colliding beams could be adopted.
For a nonzero signal involving sidereal variations,
$\cos{\ch}=\z\cdot\Z$ is nonzero,
and $\z$ precesses about $\Z$ with 
the Earth's sidereal frequency $\Om$.
A complete map between the two bases is 
given by Eq.\ (16) of Ref.\ \cite{kla}.
For convenience in what follows,
take $\th$ and $\ph$ to be conventional polar coordinates
defined about the $\z$ axis in the laboratory frame.
If the $\z$ axis is chosen along the axis of a detector,
then $\th$, $\ph$ are the usual detector polar coordinates.

Any coefficient $\vec a$ for Lorentz violation
with laboratory-frame components $(a^1, a^2, a^3)$
has nonrotating-frame components $(a^X, a^Y, a^Z)$
given by Eq.\ (12) of Ref.\ \cite{ak2}.
This relation determines the sidereal variation of $\De \vec a$ and,
using Eq.\ \rf{dem}, of $\De\La$.
The complete momentum and sidereal-time dependence 
of the parameter $\xi$ for CPT violation
in any of the $P$ systems can then be obtained.
Noting that the laboratory-frame 3-velocity of a $P$ meson
has the form
$\vec\be = \be (\sin\th\cos\ph, \sin\th\sin\ph, \cos\th)$,
and the momentum magnitude is $p \equiv |\vec p| =\be m_P \ga(p)$,
where $\ga(p) = \sqrt{1 + p^2/m_P^2}$ as usual,
the expression for $\xi$ is found to be 
\bea
\xi &\equiv &
\xi(\hat t, \vec p) \equiv \xi(\hat t, p, \th, \ph) 
\nonumber\\
&=& 
\fr 
{\ga( p)}
{\De \la} 
\bigl\{
\De a_0 
+ \be \De a_Z 
(\cos\th\cos\ch - \sin\th \cos\ph\sin\ch)
\nonumber\\
&&
\qquad
+\be \bigl[
\De a_Y (\cos\th\sin\ch 
+ \sin\th\cos\ph\cos\ch )
\nonumber\\
&&
\qquad \qquad
-\De a_X \sin\th\sin\ph 
\bigr] \sin\Om \hat t
\nonumber\\
&&
\qquad
+\be \bigl[
\De a_X (\cos\th\sin\ch 
+ \sin\th\cos\ph\cos\ch )
\nonumber\\
&&
\qquad\qquad
+\De a_Y \sin\th\sin\ph 
\bigr] \cos\Om \hat t
\bigr\} ,
\label{xipt}
\eea
where $\hat t$ denotes the sidereal time.

In deriving Eq.\ \rf{xipt},
only leading-order terms in $a_\mu$ have been kept
but no other assumption about the size of $\xi$ has been made.
The result \rf{xipt} is therefore a generalization 
of Eq.\ (13) in Ref.\ \cite{ak2},
which was obtained for the $K$ system under the assumption
of small $\de_K$.
In particular,
Eq.\ \rf{xipt} holds for the heavy-meson systems
where the possibility of large $|\xi |\gsim 1$  
remains experimentally admissible at present.

Note that the expressions \rf{dem} and \rf{xipt} 
explicitly show that the real and imaginary parts
of $\xi$ are connected through the mass and lifetime differences
of the two physical eigenstates $P_a$, $P_b$
\cite{kp}.
The relationship is
\beq
\Re\xi = - 2 \De m \Im\xi/\De \ga.
\label{reim}
\eeq
However,
in the interest of generality this result
is used only sparingly in this work. 

\section{Experiment}
\label{expt}

To illustrate some implications of the result \rf{xipt},
this section derives 
some experimentally relevant decay amplitudes, 
probabilities,
and asymmetries.
For simplicity,
attention is restricted to the case of semileptonic decays
into a final state $f$ or its conjugate state $\overline{f}$.
Although studying these decays suffices for present purposes,
other decays are also likely to be relevant in practice,
and it would be of interest to perform a more complete study. 
Another simplification adopted here is the neglect of
any violations of the 
$\De Q = \De S$,
$\De Q = \De C$,
or $\De Q = \De B$ rules.
A careful consideration of these and other more mundane complications 
would certainly be important in a definitive experimental analysis
\cite{lsx}.
However,
since there is no reason to expect such complications to exhibit 
observable momentum or sidereal-time dependences,
the extraction of a compelling positive signal 
for CPT violation should be feasible.

Under these assumptions,
the basic transition amplitudes for semileptonic decays
can be taken as
\bea
\bra{f}T\ket{P^0}
= F , &
\qquad 
\bra{f}T\ket{\overline{P^0}}
= 0,
\nonumber \\
\bra {\overline f}T\ket {\overline{P^0}}
= \overline F, &
\qquad 
\bra {\overline f}T\ket{P^0}
= 0 .
\label{amp}
\eea
Note that this parametrization allows for direct CPT violation,
which is proportional to the difference
$F^* - \overline F$,
as well as direct T violation.

To determine the time-dependent decay amplitudes and probabilities,
it is useful to obtain an explicit expression for the
time evolution of the neutral-$P$ states.
The wave functions 
$\ket{P^0}$ and $\ket{\overline{P^0}}$
can be constructed in terms of $\ket{P_a}$ and $\ket{P_b}$,
and their evolution with the meson proper time $t$ 
can then be incorporated 
via Eq.\ \rf{timevol}.
This gives
\beq
\left( \begin{array}{lr}
P^0(t, \hat t, \vec p)
\\ 
\overline{P^0}(t,\hat t,\vec p)
\end{array}
\right)
=
\left( \begin{array}{lr}
C + S \xi 
& 
S VW
\\
S VW^{-1} 
& 
C - S \xi 
\end{array}
\right)
\left( \begin{array}{lr}
P^0
\\ 
\overline{P^0}
\end{array}
\right).
\label{matrixtimevol}
\eeq
The functions $C$ and $S$ depend 
on the meson proper time $t$ and are given by
\bea
C &=&
\cos (\half \De\la t)
\exp(-\half i \la t)
\nonumber\\
&=&
\half(e^{-i\la_at} + e^{-i\la_bt})
\nonumber\\
S &=&
- i \sin (\half \De\la t)
\exp(-\half i \la t)
\nonumber\\
&=&
\half(e^{-i\la_at} - e^{-i\la_bt}).
\eea
In addition to the proper-time dependence in $S$ and $C$,
Eq.\ \rf{matrixtimevol} also contains 
sidereal time and momentum dependence from $\xi(\hat t, \vec p)$.
Since the meson decays occur quickly on the scale of sidereal time,
it is an excellent approximation to treat sidereal time $\hat t$
as a parameter independent of the meson proper time $t$.
It is therefore appropriate to take $\xi$ 
as independent of $t$
but varying with $\hat t$.
This approximation is implemented in what follows.

\subsection{Uncorrelated Mesons}
\label{uncorr}

For the case of uncorrelated meson decays,
the time-dependent decay probabilities can be obtained by 
combining Eqs.\ \rf{matrixtimevol} and \rf{amp}.
This gives 
\bea
P_f(t,\hat t,\vec p)
& \equiv &
|\bra{f}T\ket{P(t),\hat t,\vec p}|^2
\nonumber\\
&&
\hbox{\hskip-40pt}
=
\frac 1 2
|F|^2
e^{-\ga t/2} 
\nonumber\\
&&
\hbox{\hskip-20pt}
\quad
\times
\bigl[ 
(1 + |\xi|^2) \cosh\De\ga t/2
+(1 - |\xi|^2) \cos\De m t
\nonumber\\
&&
\qquad
\qquad
\hbox{\hskip-20pt}
- 2 \Re \xi \sinh\De\ga t/2
- 2 \Im \xi \sin \De m t
\bigr],
\nonumber\\        
\overline{P}_{\overline f} (t,\hat t,\vec p)
& \equiv &
|\bra{\overline f}T\ket{\overline P(t,\hat t,\vec p)}|^2
=P_f(\xi \rightarrow -\xi,
F \rightarrow \overline{F}),
\nonumber\\ 
P_{\overline f} (t,\hat t,\vec p)
& \equiv &
|\bra{\overline f}T\ket{P(t,\hat t,\vec p)}|^2
\nonumber\\
&&
\hbox{\hskip-40pt}
=
\frac 1 2 
|\overline{F}|^2
w^2 |1 - \xi^2| ~ e^{-\ga t/2}
( \cosh \De \ga t/2 - \cos \De m t),
\nonumber\\        
\overline{P}_f(t,\hat t,\vec p)
& \equiv &
|\bra{f}T\ket{\overline P(t,\hat t,\vec p)}|^2
=P_{\overline f}
(w \rightarrow 1/w, \overline{F} \rightarrow F),
\nonumber\\
\label{prob}
\eea        
where the dependence on sidereal time $\hat t$ and momentum $\vec p$
is inherited from that of $\xi$ in Eq.\ \rf{xipt}.
Inspection of these equations reveals that nonzero indirect CPT violation 
changes the shape of the first two probabilities,
while both CPT and T violation merely scale the latter two.
I emphasize that these expressions are valid
for CPT and T violation of arbitrary size.
They are also manifestly independent of the choice of phase convention
\cite{fn2}.

To extract the CPT and T violation 
from the decay probabilities \rf{prob},
it is useful to construct appropriate asymmetries.
For the case of T violation,
the dependence on sidereal time and meson momentum
has relatively little effect.
The last two probabilities in Eq.\ \rf{prob}
have the same CPT but different T dependences,
and their difference divided by their sum
is sensitive to the parameter $w$ for T violation
but independent of the parameter $\xi$ for CPT violation
and hence independent of sidereal time and meson momentum.
In contrast,
for the case of CPT violation
the situation is more involved
and several new features appear.

As a simple example illustrating some of the effects,
consider the case where $F^* = \ol F$,
i.e., neglible direct CPT violation.
The usual procedure is to assume constant nonzero $\xi$
(which is inconsistent with quantum field theory, as discussed above)
and define an asymmetry $\cA_{\rm CPT}(t)$ for CPT violation as
\beq
\cA_{\rm CPT}(t) = 
\fr{
\overline{P}_{\overline{f}}(t) - P_f(t) 
}{
\overline{P}_{\overline{f}}(t) + P_f(t) 
}.
\label{usualasymm}
\eeq
The comparable definition in the present context is still useful 
but results in an asymmetry depending also on sidereal time 
and meson momentum:
\bea
\cA_{\rm CPT}(t,\hat t,\vec p) &\equiv& 
\fr{
\overline{P}_{\overline{f}}(t,\hat t,\vec p) - P_f(t,\hat t,\vec p) 
}{
\overline{P}_{\overline{f}}(t,\hat t,\vec p) + P_f(t,\hat t,\vec p) 
}
\nonumber\\
&&
\hbox{\hskip-50pt}
=\fr{
 2 \Re \xi \sinh\De\ga t/2
+ 2 \Im \xi \sin \De m t
}{
(1 + |\xi|^2) \cosh\De\ga t/2
+(1 - |\xi|^2) \cos\De m t
},
\label{corrasymm}
\eea
where the $\hat t$, $\vec p$ dependence of $\xi$ is understood.

In practice,
the efficient practical application of this 
and related asymmetries depends on the nature of the experiment.
Appropriate averaging over one of more of the variables
$t$, $\hat t$, $p$, $\th$, $\ph$
either before or after constructing the asymmetry \rf{corrasymm}
can aid the clean extraction of bounds on $\De a_\mu$.
For instance,
under certain circumstances it may be useful to
sum the data over $\ph$ and use an asymmetry like Eq.\ \rf{corrasymm} 
but defined with the $\ph$-average of Eq.\ \rf{prob}.
The form of Eq.\ \rf{xipt} shows that binning
the data in $\hat t$ typically provides information 
on $\De a_X$ and $\De a_Y$,
while binning in $\th$ permits the separation of the spatial and
timelike components of $\De a_\mu$.
The $p$ dependence can also be useful
\cite{ak,ak2}.

As a specific example,
already used in the $K$ system
\cite{ak2,k99},
suppose the mesons involved are highly collimated 
in the laboratory frame.
Then,
the 3-velocity can be written $\vec\be = (0,0,\be )$
and the expression \rf{xipt} for $\xi$ simplifies to 
\bea
\xi (\hat t, \vec p) &=& 
\fr {\ga} {\De\la}
[ \De a_0 + \be \De a_Z \cos\ch 
\nonumber\\
&& + \be \sin\ch ( \De a_Y \sin\Om \hat t + \De a_X \cos\Om \hat t ) ] .
\label{deptcoll}
\eea
Binning in $\hat t$ therefore provides sensitivity to 
the equatorial components $\De a_X$, $\De a_Y$,
while averaging over $\hat t$ eliminates them altogether.
Indeed, 
a conventional measurement 
that ignores the dependence on sidereal time and meson momentum
is typically sensitive only to the average magnitude
\beq
|\overline {\xi}| = 
{\overline{\ga}} 
| \De a_0 + \overline{\be} \De a_Z \cos\ch |/
{|\De \la|} 
\quad ,
\label{deptcollav2}
\eeq
where 
$\overline{\be}$ and $\overline{\ga}$ 
are averages weighted over the meson-momentum spectrum.
This shows explicitly that previous analyses 
performed under the assumption of constant CPT parameter
produce results dependent on the type of experiment.

If CPT violation is small so $\xi < 1$,
the asymmetry \rf{corrasymm} takes the form
\beq
\cA_{\rm CPT}(t,\hat t,\vec p) 
\approx
\fr{
 2 \Re \xi \sinh\De\ga t/2
+ 2 \Im \xi \sin \De m t
}{
\cosh\De\ga t/2 + \cos\De m t
}.
\eeq
A further assumption that could be countenanced involves
the approximation of small $\De\ga t/2$,
i.e., $t < 2/\De\ga$.
This gives
\beq
\cA_{\rm CPT}(t,\hat t,\vec p) \approx 
\fr{
 \Re \xi \De\ga t
+ 2 \Im \xi \sin \De m t
}{
1 + \cos\De m t
}.
\eeq
It is tempting also to neglect as small the term involving $\Re\xi$,
but this is potentially invalid because $\Re\xi \propto \Im\xi/\De\ga$
according to Eq.\ \rf{reim}.
Imposing the prediction \rf{reim} instead gives
\beq
\cA_{\rm CPT}(t,\hat t,\vec p) \approx
\fr{
 2 \Im \xi ( \sin \De m t - \De m t )
}{
1 + \cos\De m t
}.
\eeq

The extraction of complete information about $\De a_\mu$
requires clean CPT tests involving asymmetries such as 
Eq.\ \rf{corrasymm} that are independent of the parameter $w$
for T violation.
However,
the dependence on sidereal time of certain CPT-violating effects 
offers the possibility of extracting clean CPT bounds
on spatial components of $\De a_\mu$
even using observables that mix T and CPT effects
\cite{ak2}.
An example is provided by the standard rate asymmetry $\de_l$
for $K_L$ semileptonic decays 
\cite{pdg}:
\beq
\de_l \equiv
\fr{\Ga (K_L \to l^+\pi^-\nu)
- \Ga(K_L \to l^-\pi^+\overline{\nu})}
{\Ga (K_L \to l^+\pi^-\nu)
+ \Ga(K_L \to l^-\pi^+\overline{\nu})} ,
\label{dell}
\eeq
which under the assumption 
of constant nonzero parameter for CPT violation 
(inconsistent with quantum field theory, as noted above) 
is determined by a combination of T and CPT effects
that cannot be disentangled without further information.
In the $w\xi$ formalism,
the asymmetry \rf{dell} and its generalization to
arbitrary $P_b$ is found to be 
\bea
\de_l (\hat t, \vec p) &\equiv &
\fr
{\Ga (P_b \to f)
- \Ga(P_b \to \overline{f})}
{\Ga (P_b \to f)
+ \Ga(P_b \to \overline{f})}
\nonumber \\
&=&
\fr
{|1-\xi^2| - |1 + \xi|^2 w^2}
{|1-\xi^2| + |1 + \xi|^2 w^2}
\nonumber \\
&\approx &
(1 - w) - \Re \xi (\hat t, \vec p)
\label{dellhatt}
\eea
where the last line assumes $w \approx 1$, $\xi \ll 1$,
i.e., small T and CPT violation.
Binning in sidereal time or momentum 
can therefore under suitable circumstances
bound the spatial components of $\De a_\mu$ 
independently of T violation,
even for observables involving both T and CPT violation.

\subsection{Correlated Mesons}
\label{corr}

Another situation of experimental importance is 
the case of correlated meson pairs,
resulting from quarkonium production and decay.
The normalized initial quantum state
ensuing immediately after the strong decay of the quarkonium  
can be written as
\beq
\ket{i} = 
\fr 1{\sqrt{2}}
\bigl(
\ket{P^0(+)} \ket{\overline{P^0}(-)}
-\ket{P^0(-)} \ket{\overline{P^0}(+)}
\bigr),
\eeq
where $(+)$ indicates the meson travels 
in a specified direction in the quarkonium rest frame
while $(-)$ indicates it travels in the opposite direction.
Note that this initial state is independent 
of the choice of phase convention.

Let the meson moving in the $(+)$ direction
have 3-momentum $\vec p_1$ in the laboratory frame
and decay into a final state $f_1$ at proper time $t_1$.
Similarly,
let the other meson have 3-momentum $\vec p_2$
and decay into a final state $f_2$ at proper time $t_2$.
As before,
in tracking the sidereal-time dependence,
it is an excellent approximation to regard
the time interval between quarkonium production 
and detection of the decay products as negligible
on the scale of the Earth's rotation period,
so in what follows the creation of the state $\ket{i}$ 
and its evolution through the double decay process
are taken to occur at fixed sidereal time $\hat t$.

The probability amplitude $A_{f_1f_2}$
for the double decay 
can be regarded as a function
of the decay times $t_1$, $t_2$,
of the sidereal time $\hat t$,
and of the two meson momenta $\vec p_1$, $\vec p_2$.
It is given by
\bea
A_{f_1f_2}
&\equiv&
A_{f_1f_2}(t_1,t_2,\hat t,\vec p_1,\vec p_2)
=
\bra{f_1 f_2}T\ket{i} 
\nonumber\\
&=&
\fr 1{\sqrt{2}}
\bigl[
\bra{f_1}T\ket{P^0(t_1,\hat t,\vec p_1)}
\bra{f_2}T\ket{\overline{P^0}(t_2,\hat t,\vec p_2)}
\nonumber\\
&&
\quad
- \bra{f_1}T\ket{\overline{P^0}(t_1,\hat t,\vec p_1)}
\bra{f_2}T\ket{P^0(t_2,\hat t,\vec p_2)}
\bigr].
\label{amp12}
\eea
The time evolutions of 
$\ket{P^0(t,\hat t,\vec p)}$
and $\ket{\overline{P^0}(t, \hat t, \vec p)}$
are determined by
Eq.\ \rf{matrixtimevol}.
In substituting these expressions 
into the decay amplitude \rf{amp12},
care is required to keep separate track
of the CPT-violating parameters $\xi_1$ and $\xi_2$
for each meson,
since they depend on the meson 3-momenta
and therefore typically differ 
in accordance with Eq.\ \rf{xipt}.

It is convenient and feasible to write a single expression
that holds for all double decay modes,
including the various double-semileptonic combinations.
For $a = 1,2$,
define
\beq
\bra{f_a}T\ket{P^0}
= F_a , 
\qquad 
\bra{f_a}T\ket{\overline{P^0}}
= \overline F_{a}, 
\label{amps}
\eeq
and let $C_a = C(t_a)$, $S_a = S(t_a)$.
Then, 
the probability amplitude is found to be
\bea
A_{f_1f_2} &=& 
\fr 1{\sqrt{2}}
\bigl[
(F_1\ol F_2 + F_2\ol F_1)
(\xi_1S_1C_2 - \xi_2S_2C_1)
\nonumber\\
&&
\hbox{\hskip-20pt}
+ (F_1\ol F_2 - F_2\ol F_1)
(C_1C_2 - (\xi_1\xi_2+V_1V_2)S_1S_2)
\nonumber\\
&&
\hbox{\hskip-20pt}
+ (F_1 F_2 W^{-1} - \ol F_1\ol F_2 W)
(V_2 C_1 S_2 - V_1 S_1 C_2)
\nonumber\\
&&
\hbox{\hskip-20pt}
+ (F_1 F_2 W^{-1} + \ol F_1\ol F_2 W)
(\xi_1 V_2 - \xi_2 V_1) S_1 S_2
\bigr],
\eea
where the dependence on $\hat t$ and $\vec p_1$, $\vec p_2$
is understood.
The quantities $V_1$, $V_2$ are defined in terms of $\xi_1$, $\xi_2$
by Eq.\ \rf{uvdef},
while $W= w\exp(i\om)$ as before.

Next,
consider the special case of double-semileptonic decays
and adopt the notation of Eq.\ \rf{amp}.
It is useful to introduce the definitions
\beq
t = t_1 + t_2,
\quad
\De t = t_1 - t_2.
\eeq
In terms of these variables,
some algebra yields the four possible decay amplitudes as
\bea
A_{f\ol f}
&=&
\fr {F\ol F}{ 2 \sqrt{2}}
\bigl[
(1 - \xi_1\xi_2 - V_1V_2)
\cos \half \De\la t
\nonumber\\
&&
\qquad
+
(1 + \xi_1\xi_2 + V_1V_2)
\cos \half \De\la\De t
\nonumber\\
&&
\qquad
- i (\xi_1 - \xi_2)
\sin \half \De\la t
\nonumber\\
&&
\qquad
- i (\xi_1 + \xi_2)
\sin \half \De\la \De t
\bigr]
~ e^{-i\la t/2},
\nonumber\\
A_{\ol ff}
&=&
-A_{f\ol f}(\xi_1 \rightarrow -\xi_1, \xi_2 \rightarrow -\xi_2),
\nonumber\\
A_{ff}
&=&
\fr {F^2}{ 2 \sqrt{2}}
W^{-1}\bigl[
(\xi_1V_2 - \xi_2V_1)
(\cos \half \De\la t
-\cos \half \De\la\De t)
\nonumber\\
&&
\qquad
\qquad
+ i (V_1 - V_2)
\sin \half \De\la t
\nonumber\\
&&
\qquad
\qquad
+ i (V_1 + V_2)
\sin \half \De\la \De t
\bigr]
~ e^{-i\la t/2},
\nonumber\\
A_{\ol f\ol f}
&=&
-A_{ff}(F\rightarrow \ol F,
W \rightarrow W^{-1},
\xi_1 \rightarrow -\xi_1, \xi_2 \rightarrow -\xi_2),
\nonumber\\
\label{corramp}
\eea
where the dependence on $\hat t$ and $\vec p_1$, $\vec p_2$
is again understood.

The expressions \rf{corramp} 
are valid for CPT and T violation of arbitrary size
and are independent of phase conventions.
Nontrivial sensitivity to
the sum and difference of $\xi_1$ and $\xi_2$ is manifest.
The corresponding decay probabilities are 
straightforward to obtain but are somewhat cumbersome.
They inherit the independence of phase conventions
and the nontrivial sensitivity to $\xi_1 \pm \xi_2$.
Since these factors depend on all four parameters $\De a_\mu$
for CPT violation,
appropriate analysis of experimental data for correlated decays 
can provide four independent CPT tests.

The type of analysis needed depends on the experimental situation.
The remarks following Eq.\ \rf{corrasymm} about averaging
and binning apply here,
and there are also considerations specific to the case
of correlated mesons.
For example, 
if the quarkonium is produced at rest in the laboratory,
perhaps by a symmetric collider,
then the 3-momenta of the correlated mesons are  
equal in magnitude and opposite in direction.
The sum 
\beq
\xi_1 + \xi_2 = 2 \ga (p)\De a_0/\De\la
\label{sum}
\eeq
is then independent of $\De \vec a$,
so extracting an asymmetry sensitive to $\xi_1 + \xi_2$ 
yields a clean bound on $\De a_0$. 
Similarly,
the difference $\xi_1 - \xi_2$ is independent of $\De a_0$,
and binning in sidereal time permits bounds on
the three components $\De \vec a$.
If instead the quarkonium is produced in an asymmetric collider,
the two 3-momenta of the correlated mesons are \it not \rm
back-to-back in the laboratory frame,
so $\xi_1 \pm \xi_2$ are both sensitive 
to all components of $\De a_\mu$.
Four independent measurements of CPT violation can again be extracted.

Many of the interesting features can be illustrated
in the approximation of small $\xi_1$, $\xi_2$,
for which the expressions simplify to some extent.
This approximation is certainly valid for the $K$ system,
and the recent results from OPAL, DELPHI, and BELLE
\cite{bexpt}
imply it is also valid for the $B_d$ system.
The situation for the $D$ and the $B_s$ systems is less clear,
with large CPT violation remaining experimentally admissible,
but many of the following considerations still apply. 

Consider for definiteness the double decay into $f\overline{f}$.
To leading order in $\xi_1$ and $\xi_2$,
the decay probability $P_{f\ol f}$ is 
\bea
P_{f\ol f} &=&
P_{f\ol f}(t, \De t, \hat t, \vec p_1, \vec p_2)
\nonumber\\
&&
\hbox{\hskip-20pt}
=
\frac 1 4 |F\ol F|^2 
~ e^{-\ga t/2}
\bigl\{
\cosh \half \De\ga \De t
+
\cos \De m\De t
\nonumber\\
&&
-\Re (\xi_1 + \xi_2)
\sinh \half \De\ga \De t
\nonumber\\
&&
-\Im(\xi_1 + \xi_2)
\sin\De m\De t
\nonumber\\
&&
+ 2\Im 
\bigl[ (\xi_1 - \xi_2)
\cos (\half \De\la^*\De t)
\sin (\half \De\la t)
\bigr]
\bigr\}.
\nonumber\\
\label{prob2}
\eea
This expression shows the combination $\xi_1 + \xi_2$
is associated with an odd function in $\De t$, 
while $\xi_1 - \xi_2$ is associated with an even function in $\De t$. 
This distinction allows the separate extraction
of $\xi_1 \pm \xi_2$.
As an explicit example,
the case of the sum $\xi_1 + \xi_2$ is treated here.

In typical experimental situations 
for the correlated double-meson decay,
the time sum $t$ is unobservable
but the difference $\De t$ can be used as a fitting parameter.
It is therefore appropriate to work with
an integrated probability 
$\Ga _{f\ol f}(\De t, \hat t, \vec p_1, \vec p_2)$
obtained by integrating the probability \rf{prob} over $t$:
\beq
\Ga _{f\ol f}(\De t, \hat t, \vec p_1, \vec p_2) = 
\int_{|\De t|}^\infty dt~
P_{f\ol f} (t, \De t, \hat t, \vec p_1, \vec p_2).
\eeq
An asymmetry $\cA_{{\rm CPT},f\ol f}$ sensitive to 
the sum $\xi_1 + \xi_2$ of parameters for CPT violation
can then be defined as 
\bea
\cA_{{\rm CPT},f\ol f} &=&
\cA_{{\rm CPT},f\ol f}(\De t, \hat t, \vec p_1, \vec p_2)
\nonumber \\
&=&
\fr{
\Ga _{f\ol f}(\De t, \hat t, \vec p_1, \vec p_2) 
- \Ga _{f\ol f}(-\De t, \hat t, \vec p_1, \vec p_2)
}{
\Ga _{f\ol f}(\De t, \hat t, \vec p_1, \vec p_2)
+ \Ga _{f\ol f}(-\De t, \hat t, \vec p_1, \vec p_2)
}.
\label{asymmfbarf}
\eea
Calculation gives
\bea
\cA_{{\rm CPT},f\ol f}=
&&
\nonumber\\
&&
\hbox{\hskip-35pt}
\fr{
-\Re (\xi_1 + \xi_2)
\sinh \half \De\ga \De t
-\Im(\xi_1 + \xi_2)
\sin\De m\De t
}{
\cosh \half \De\ga \De t
+
\cos \De m\De t
},
\nonumber\\
\label{sumxiasymm}
\eea
which is valid to lowest order in CPT-violating quantities. 
For the $B_d$ system,
this expression generalizes the asymmetry obtained 
\cite{msas}
under the assumption of constant parameter for CPT violation
and used to place the recent experimental limits 
on CPT violation at BELLE
\cite{bexpt}.

For quarkonia produced in a symmetric collider 
the asymmetry \rf{sumxiasymm} depends only on $\De a_0$
because the sum $\xi_1 + \xi_2$ is given by Eq.\ \rf{sum}.
There is therefore no variation with $\hat t$,
and the line spectrum of the mesons
implies there is also no variation with $\vec p_1 = - \vec p_2$.
In this case,
a direct fit to the variation with $\De t$
provides a bound on $\De a_0$.

In contrast,
for quarkonia produced in an asymmetric collider
the asymmetry \rf{sumxiasymm} depends on all four parameters $\De a_\mu$
and also varies with $\hat t$ and $\vec p_1$, $\vec p_2$. 
For any given situation,
forming an asymmetry of the type \rf{asymmfbarf}
after averaging Eq.\ \rf{prob2}
over suitable combinations of the 
variables $\hat t$, $\vec p_1$, $\vec p_2$
permits the extraction of four independent CPT bounds,
one for each parameter $\De a_\mu$.  
Independent tests of this kind for the $B_d$ system 
should be feasible at both BaBar and BELLE,
where the quarkonia are produced in asymmetric collisions
and the meson pairs are boosted in the laboratory frame.

\section{Summary}

This work has studied some aspects of tests of CPT and Lorentz symmetry 
using neutral-meson oscillations.
A formalism has been adopted 
for the treatment of arbitrarily large 
indirect CPT and T violation
in the $K$, $D$, $B_d$, and $B_s$ systems
that is phase-convention independent.
It involves a real parameter $w$ for T violation
and a complex parameter $\xi$ for CPT violation.
An expression for the latter,
given as Eq.\ \rf{xipt},
is derived from the general Lorentz- 
and CPT-breaking standard-model extension.
This equation reveals that CPT observables can vary
with the magnitude and orientation of the meson momentum
and hence also with sidereal time.
To illustrate some of the implications for experiment,
transition amplitudes, decay probabilities, 
and sample CPT-sensitive asymmetries
for semileptonic decays are derived.
Both uncorrelated and correlated mesons are considered,
and some consequences for experiments are described.

The analysis shows that four independent experimental bounds
are required to bound CPT violation completely in any single 
neutral-meson system.
Since these parameters may differ between systems,
separate experimental analyses are required in each case.
No bounds are available in the $D$ or $B_s$ systems as yet.
Certain combinations of the four key parameters $\De a_\mu$
have been constrained in the $K$ and $B_d$ systems
by recent experiments
\cite{k99,bexpt},
but no definitive analysis has yet been performed.
Obtaining a complete set of linearly independent measurements 
in any of the meson systems 
has the potential to offer our first glimpse of
physics at the Planck scale
and would in any case provide crucial experimental information 
on the existence of CPT and Lorentz violation in nature.

\section*{Acknowledgments}

This work was supported in part 
by the United States Department of Energy 
under grant number DE-FG02-91ER40661.

\begin{appendix}
\section{Standard Formalisms}

This appendix lists a few key properties of 
five standard formalisms for indirect T and CPT violation.
All these can be traced to early work 
several decades ago in the context of the $K$ system
\cite{lw}.
For most of these standard formalisms,
several closely related variants exist in the literature,
but for definiteness only one of each type is presented here.

The $M\Ga$ formalism sets
\beq
\La = M - \half i \Ga =
\left( \begin{array}{lr}
M_{11} - \half i \Ga_{11}
& 
M_{12} - \half i \Ga_{12}
\\ & \\
M_{12}^* - \half i \Ga_{12}^*
& 
M_{22} - \half i \Ga_{22}
\end{array}
\right).
\label{lm}
\eeq
The off-diagonal quantities are all phase-convention dependent.
The parameter for CPT violation is the combination
$(M_{11} - M_{22}) - i (\Ga_{11} - \Ga_{22})/2$.
The parameter for T violation is
$|( M_{12}^* - i \Ga_{12}^*/2) /(M_{12} - i \Ga_{12}/2)|$.
The masses and decay rates are given by
\bea
\la &=&  (M_{11} + M_{22}) - \half i (\Ga_{11} + \Ga_{22}),
\nonumber\\
\De\la &=& 
2\bigl[ 
(M_{12} - \half i \Ga_{12}) 
(M^*_{12} - \half i \Ga^*_{12}) 
\nonumber\\
&&
+\frac 14[ (M_{11} - M_{22}) - \half i (\Ga_{11} - \Ga_{22})]^2
\bigr]^{1/2},
\label{mgamess}
\eea
where the definitions in Eq.\ \rf{ldl} are understood to hold.

The $D E_1 E_2 E_3$ formalism sets
\beq
\La = 
\left( \begin{array}{lr}
-i D + E_3 
&
E_1 - i E_2
\\ & \\
E_1 + i E_2 
&
-i D - E_3
\end{array}
\right).
\label{de123}
\eeq
All off-diagonal quantities are phase-convention dependent.
The parameter for CPT violation is $E_3$.
The parameter for T violation is
$i(E_1 E_2^* - E_1^* E_2)$.
The masses and decay rates are given by 
$\la = - 2 i D$,
$\De \la = 2 \sqrt{E_1^2 + E_2^2 + E_3^2}$.

The $D E \th \ph$ formalism sets
\beq
\La = 
\left( \begin{array}{lr}
-i D + E \cos \th 
&
E \sin \th e^{-i\ph}
\\ & \\
E \sin \th e^{i\ph}
&
-i D - E \cos \th 
\end{array}
\right).
\label{dethph}
\eeq
The parameter $\ph$ is phase-convention dependent.
The parameter for CPT violation is $\cos\th$.
The parameter for T violation is
$|\exp(i\ph)|$.
The masses and decay rates are given by 
$\la = - 2 i D$,
$\De \la = 2 E$.

There are also formalisms that are introduced 
in terms of the relationship 
between the strong-interaction eigenstates
$P^0$, $\overline{P^0}$
and the physical eigenstates
$P_a$, $P_b$.
A general one is the $pqrs$ formalism,
which sets
\bea
\ket{P_a} &=&
p \ket{P^0} + q \ket{\overline{P^0}} , 
\nonumber \\
\ket{P_b} &=&
r \ket{P^0} -s \ket{\overline{P^0}} ,
\label{pqrsdef}
\eea         
where $p$, $q$, $r$, $s$ are complex parameters.
In this formalism,
one can show
\beq
\La = 
\fr 1 {2(ps+qr)}
\left( \begin{array}{lr}
\la (ps+qr) 
&
2\De\la pr
\\
+\De\la (ps-qr)
&
\\ & \\
&
\la (ps+qr) ~~\quad
\\
2\De\la qs
&
- \De\la (ps-qr)
\end{array}
\right).
\label{pqrs}
\eeq
The complex parameters $p$, $q$, $r$, $s$ 
are all phase-convention dependent.
They are also substantially redundant, 
since only three of their eight real components have physical meaning.
The normalization conventions for the wave functions
represent two degrees of freedom,
often fixed by the choice 
$|p|^2 + |q|^2 = |r|^2 + |s|^2 = 1$.
The remaining three unobservable degrees of freedom
are the absolute phases of $\ket{P_a}$ and $\ket{P_b}$
and the relative phase of $\ket{P^0}$ and $\ket{\overline{P^0}}$.
The parameter for CPT violation is $(ps-qr)$.
The parameter for T violation is $|pr/qs|$.
The masses and decay rates are additional independent quantities, 
taken here as  $\la $, $\De\la$.

The $\ep\de$ formalism
\cite{fn3}
is widely adopted for the $K$ system.
It can be regarded as a special case of the $pqrs$ formalism.
For arbitrary-size T and CPT violation,
the $\ep\de$ formalism can be defined as
\bea
\ket{P_a} &=&
\fr{ (1 + \ep + \de) \ket{P^0}
+(1 - \ep - \de) \ket{\overline{P^0}}  }
{ \sqrt{2( 1 + |\ep + \de|^2)}  } ,
\nonumber \\
\ket{P_b} &=&
\fr{ (1 + \ep - \de) \ket{P^0}
-(1 - \ep + \de) \ket{\overline{P^0}}  }
{ \sqrt{2( 1 + |\ep - \de|^2)}  } .
\label{epdedef}
\eea         
In this formalism,
$\La$ is given by Eq.\ \rf{pqrs}
with appropriate substitutions 
for the parameters $p$, $q$, $r$, $s$ in terms of $\ep$, $\de$,
obtained from Eq.\ \rf{epdedef}.
Both $\ep$ and $\de$ depend on phase conventions.
Nonzero values of $\ep$ and $\de$ characterize T and CPT violation,
respectively.
For the special case of small $\ep$ and $\de$,
which is a good approximation in the $K$ system,
one can show
\beq
\La \approx 
\half \left( \begin{array}{lr}
\la + 2\De\la \de
&
\De\la (1 + 2 \ep) 
\\ & \\
\De\la (1 - 2 \ep) 
&
\la - 2\De\la \de
\end{array}
\right).
\label{epde}
\eeq
Even within this approximation
$\ep$ is phase-convention dependent,
although $\de$ is not.
The parameter for T violation can then be taken to be $\Re\ep$, 
for example.
The masses and decay rates are independent quantities
and here are specified by $\la$, $\De\la$.

\end{appendix}

\end{multicols}

\begin{references}
\bibitem{fn00}
Instead of the modulus $w$,
which is positive and becomes 1 when T is preserved,
one could introduce a parameter $w^\prime \equiv \ln w$ 
that ranges over positive and negative values
and vanishes when T is preserved.
However,
expressions for certain observables then become 
cluttered with exponential factors $\exp(w^\prime)$.
\bibitem{fn0}
Note that if $\xi = \pm 1$ then
$V=0$, $\De \La = \De \la$
and there are no oscillations,
so in what follows $\xi \neq \pm 1$ and hence $V\neq 0$
is assumed.
\bibitem{fn1}
When CP is preserved, $\xi = 0$, $w = 1$,
and $\ket{P_a}$ and $\ket{P_b}$ reduce to CP eigenstates.
In a phase convention with $\om = \et_a = \et_b = 0$,
the eigenstates then take the familiar form
$\ket{P_{a,b}} = (\ket{P^0} \pm \ket{\overline{P^0}})/\sqrt 2$.
\bibitem{fn2}
Note that the transition amplitudes $F$ and $\overline{F}$ 
in Eq.\ \rf{amp} do change with the phase convention,
but the norms $|F|$, $|\overline F|$
and the product $F\overline{F}$ are invariant.
\bibitem{fn3}
Sometimes $\De$ or $-\De$ is used instead of $\de$.
\end{references}
\end{document}